# A Possible Problem with the Sunspot Number

Leif Svalgaard

## Abstract

Sunspots are areas of strong magnetic fields driven by a convective dynamo. Rudolf Wolf devised his Sunspot Number (SN) series to describe its variation with time. Most other solar phenomena vary in concert with SN, in particular the Sun's microwave radiation. During the interval 2014-Jul to 2015-Dec, the variation of the SN with the microwave flux (*e.g.* F10.7) was anomalous, although microwave flux did not exhibit any anomaly with respect to the solar magnetic field measured by the HMI instrument on the SDO spacecraft. This points to a possible problem with the derivation of the SN using the 148,714 reports of sunspot observations received by the World Data Center Solar Influences (SIDC/SILSO at ROB) since 2011 (when the last public raw data was released). The unavailability of the raw data since then, contrary to avowed 'open data policy' of ROB, prevents independent assessment and possible correction of the anomaly. We urge ROB to make the raw data available.

## Analysis

The venerable Sunspot Number is probably the most analyzed time series (with the possible exception of some stock market series) of all times. The current version (2.0 or 2.1?) really dates to 2011. By that, I mean that the raw data on which the sunspot series is based have been publicly available for confirmative research and critical analysis prior to 2011. In the years after that, this kind of transparency has been absent, and we must rely on other indicators of solar activity to track what the Sun is doing in order to assess its impact on the environment of the Earth and through the Solar System.

As sunspots are visible manifestations of magnetic fields on the Sun, measurement of solar magnetic fields [or magnetic flux over the observing aperture which is what is actually measured] gives us a more direct indicator of said activity, of which the sunspot number is merely a *proxy*. A proxy that, however, usefully stretches back more than four hundred years. Solar magnetic fields have been measured at ground-based observatories ever since George Hale in 1908 showed that sunspots were magnetic. During 1996-2011, the MDI and from 2011, the HMI instruments on the SOHO and SDO spacecrafts [Scherrer et al. 2012] have measured the solar magnetic flux over millions of pixels covering the solar disk. We measure the flux in *Maxwells*. Dividing the flux by the area of the solar disk (1.5×10$^{22}$ cm$^2$) we get a pseudo-field value which we call '*Gauss*'. Figure 1 shows monthly values of the total unsigned line-of-sight [LOS] flux over the solar disk measured by HMI (pink) and MDI (green and normalized to HMI using their values during their overlap (2010/May- 2011/April) [Svalgaard & Sun, 2016].

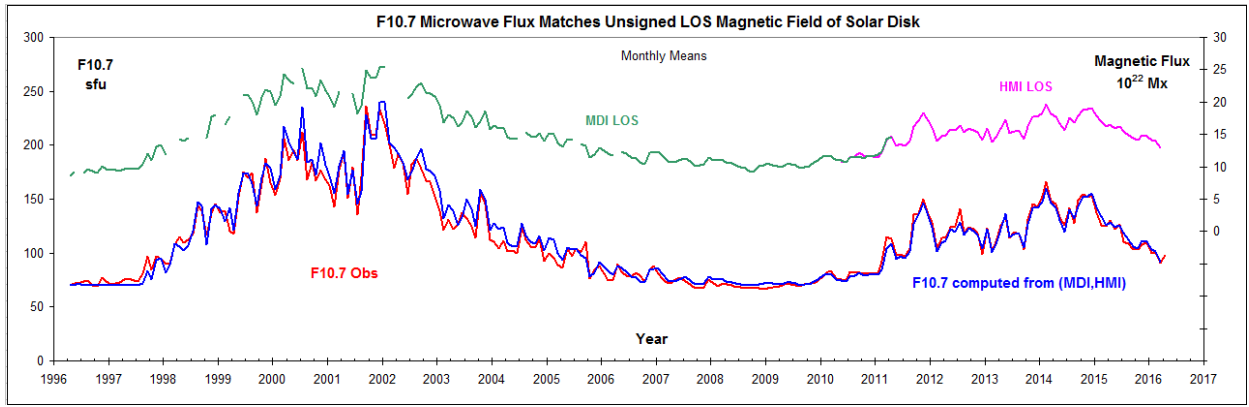

***Figure 1***: *Upper green curve: MDI\* LOS flux scaled to match the HMI LOS flux based on their overlap. Upper pink curve: HMI. Lower red curve: observed F10.7 flux reduced to 1 AU. Blue curve: F10.7\* computed from the magnetic flux of the composite record. All curves have 1-month resolution.*

There are two mechanisms that produce microwave radio emission from a non-flaring Sun: thermal bremsstrahlung TB (also called free-free emission) and gyroresonance GR. TB results from collisions between electrons and ions that depends strongly on the plasma density in the corona (that depends on heating by reconnecting magnetic fields) and the underlying chromosphere. GR emission arises from acceleration of electrons spiraling around magnetic field lines. We therefore expect the F10.7 flux to depend on the magnetic flux. The lower curves in Figure 1 bear out the expectation showing that we can calculate F10.7 from the magnetic flux [Svalgaard & Sun, 2016].

Now, nearly a decade later, we can extend the HMI record, Figure 2:

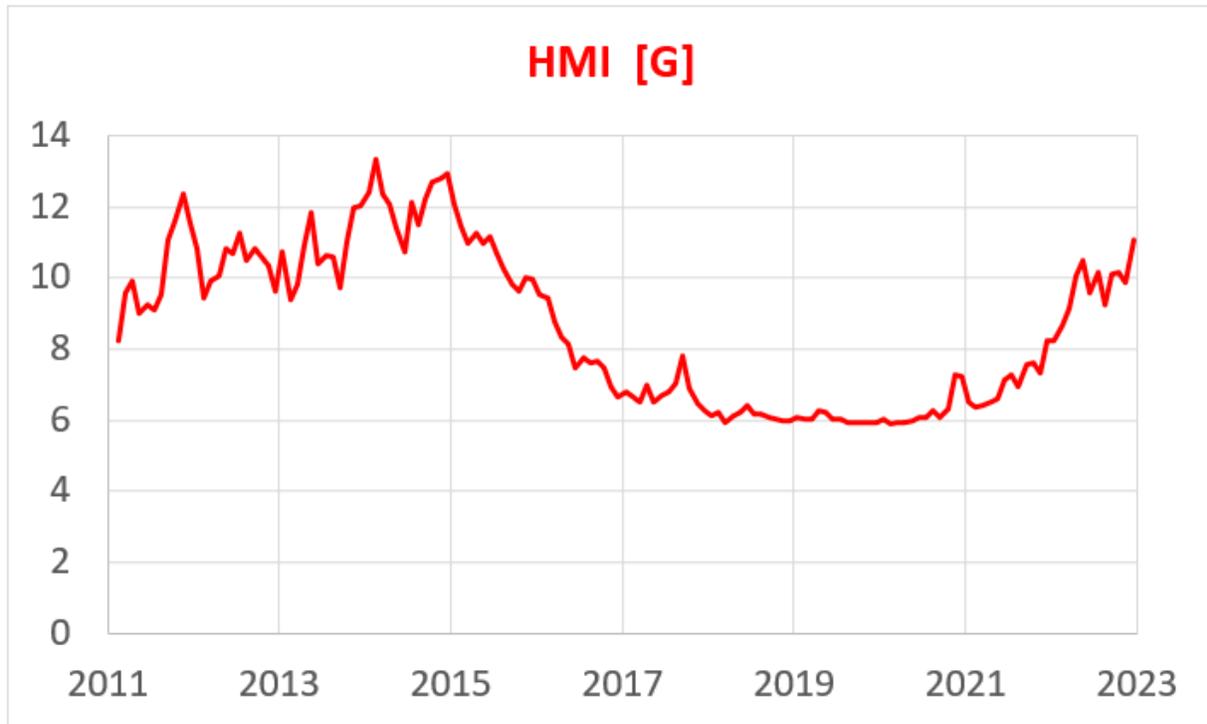

***Figure 2:*** *Red curve: Monthly values of the total disk-average unsigned pseudo-Gauss LOS from HMI (from http://jsoc.stanford.edu/). Values below 6 G include an amount of noise which we assume to be constant, i.e., not depending on solar activity.*

As before, we can correlate HMI with observed values of the microwave F10.7 flux. The relationships and results are shown in Figure 3.

It is not fully clear why there should be *two* different formulae, one before and one after 2016. The difference amounts to an upwards jump of 16%. It is known (Ken Tapping, personal communication) that early on, solar radio bursts (associated with flares) were manually removed from the F10.7 record. Currently, this is no longer done, resulting in elevation of reported fluxes (actually: flux densities).

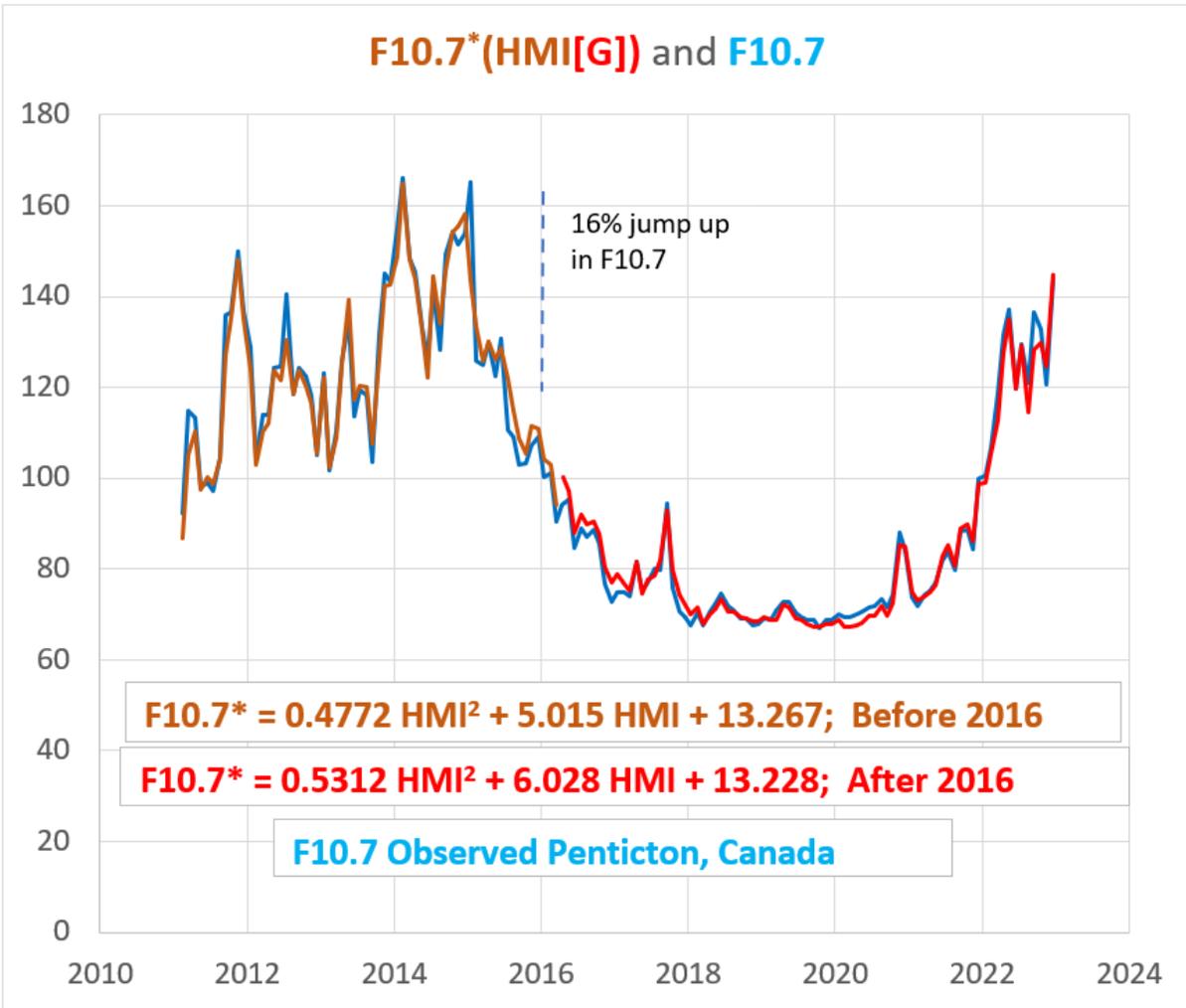

*Figure 3:* F10.7 calculated from the HMI 'field', using different formulae before and after ~2016 (shown in orange and red, respectively). The reported observed values from Penticton are shown in light blue.

At Nobeyama (Nagano Prefecture) in Japan a cluster of solar radio polarimeters have been observing at frequencies 1, 2, 3.75, and 9.4 GHz, straddling the F10.7 flux at 2800 MHz [Shimojo et al. 2024]. Continuous monitoring has been conducted for over seven decades for 3.75 GHz (although beginning at a different location). Fitting each frequency to the F10.7 flux we compute a composite representative of the 2800 MHz (F10.7) flux [for details, see Svalgaard 2010], Figure 4.

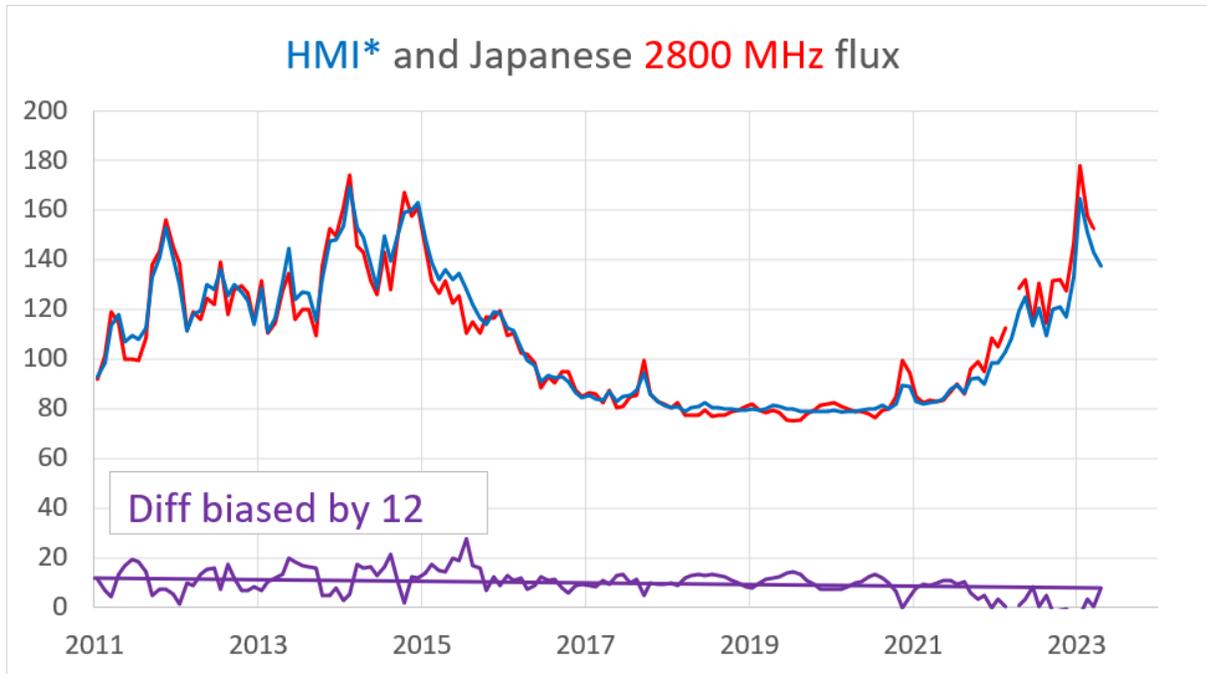

***Figure 4:*** *Monthly averages of the reconstructed 2800 MHz (F10.7) flux [red curve] and the HMI magnetic 'field' scaled [blue curve] to match the 2800 MHz flux: HMI\* = 0.7557 HMI $^2$ – 2.3056 HMI + 66.209 ($R^2$ = 0.9598). The difference between observed and calculated values is shown by the lower purple curve, biased by twelve flux units.*

The Japanese data confirms that the upward jump around 2016 of the Canadian F10.7 series is instrumental (or procedural) and that F10.7 should be adjusted as done in Figure 3. For a further check we can use the 'Bremen' Mg II index [Snow et al. 2014]. The emission cores of the Mg II doublet lines near λ280 nm show large natural solar radiation variability in the near-UV range. Observations from several spacecraft instruments in the GOME series are combined into a composite index that is compared with HMI in Figure 5.

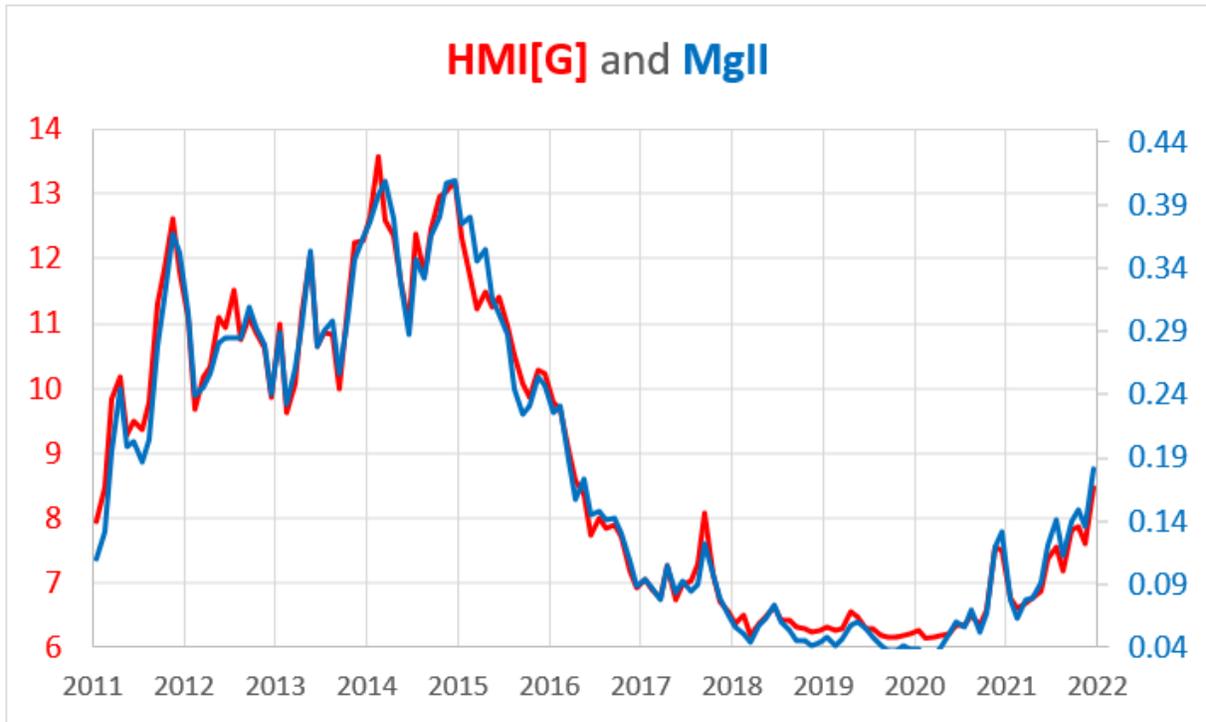

*Figure 5: Monthly averages of the HMI 'field' in G [red curve] and the Magnesium II core-to-wing ratio [blue curve] scaled to match each other. Instruments with different resolutions may give different values, but in all cases the Mg II index is the ratio of the chromospheric contribution to the photospheric contribution.*

The comparisons confirm that both the microwave data and the Mg II data show that HMI is a good indicator of solar activity. If the sunspot number, SN, is also a good indicator of solar activity (as implicitly assumed by users of the series), the variations of HMI and SN should agree (after suitable transformation). Figure 6 shows that they do except during the interval 2014/Jul – 2015/Dec (marked by the shaded box).Before the series can be compared they must first be brought on the same scale. Figure 7 shows HMI scaled to SN (version 2) for times outside of the shaded box with scaling relation HMI*=19.243HMI–110.76 ($R^2$= 0.9673).

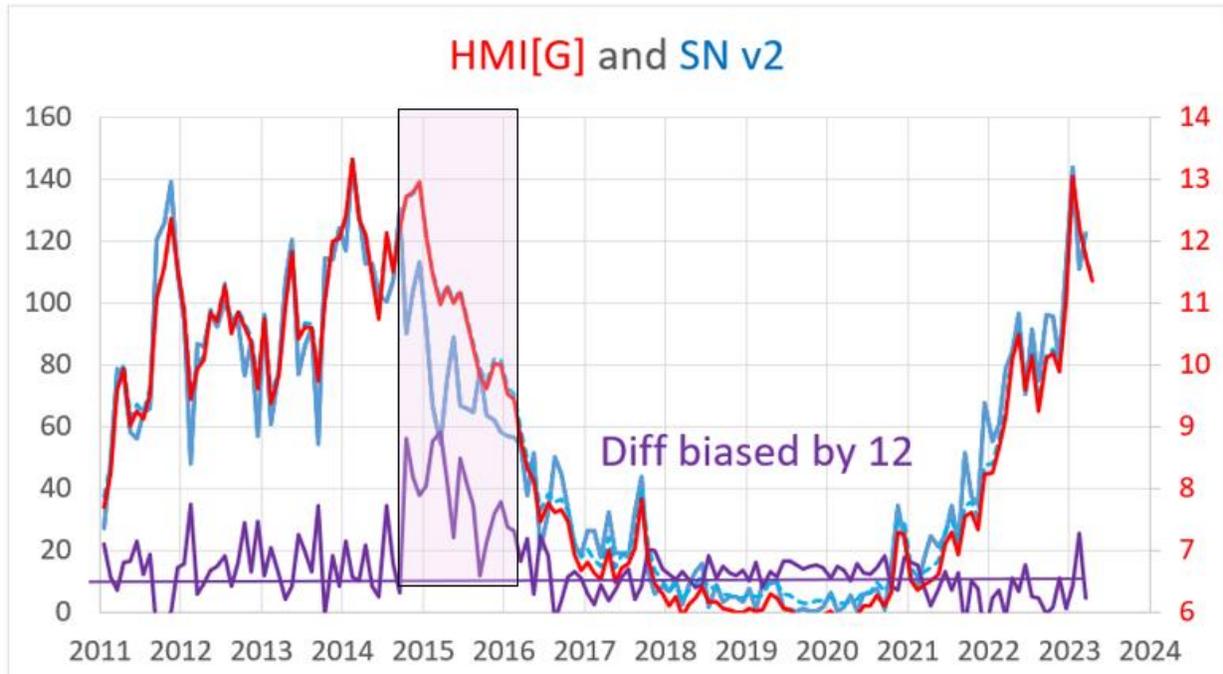

*Figure 6:* Monthly averages of the scaled HMI* [red curve] and the Sunspot Number (v2) [blue curve]. The difference HMI*-SN biased by 12 is shown by the lower [purple] curve. A region of large difference is marked by the shaded box.

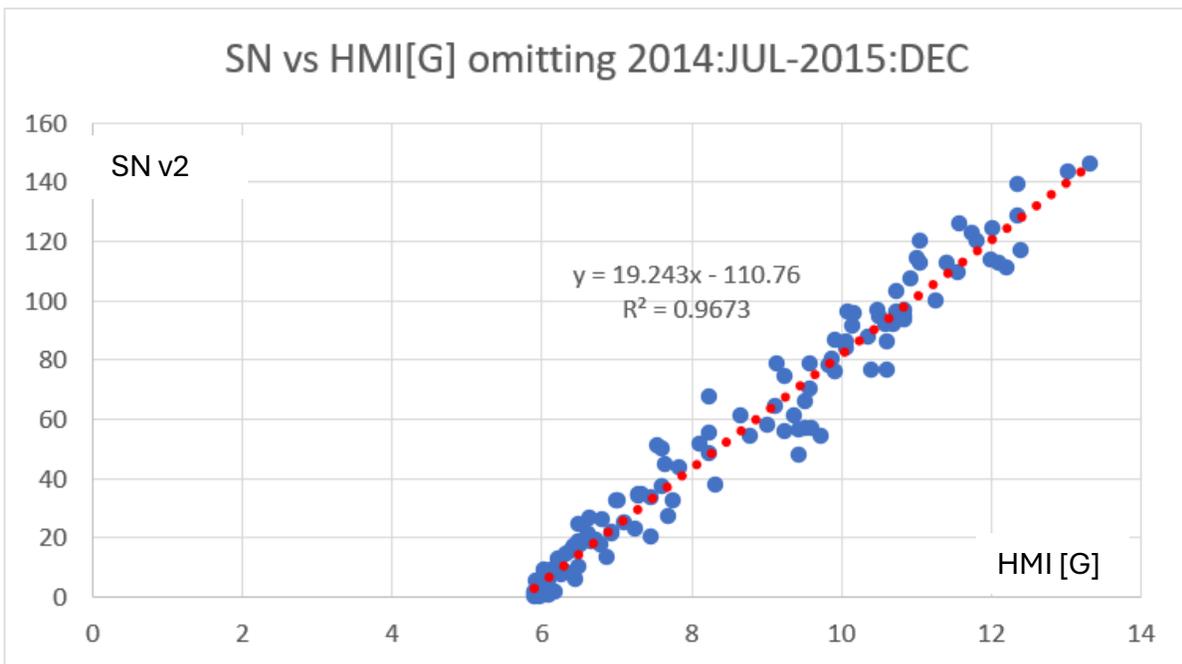

*Figure 7:* Scaling of monthly averages of HMI [G] to the Sunspot Number SN [v2] since 2011, omitting the months from 2014 July through 2015 December. The trend line [red] is linear and supports the scaling relationship indicated in the Figure.

As Clette and Lefèvre [Clette et al., 2021] remind us "inaccessibility of [the…] source data prevents researchers from access to a huge amount of detailed information and to essential metadata [… that can] feed full statistical analysis by current state-of-the-art methods and lead to an improved index" the unavailability of the raw sunspot source data held at SILSO since 2011 prevents us from researching the cause of the discrepancy shown in Figure 6.

Figure 8 shows the number of reports per year of sunspot observations received at SILSO. It is interesting that during the interval when the difference between observed and expected sunspot numbers was abnormally high (shaded box in Figure 6), the number of reported observations per year suddenly almost doubled, further emphasizing the need for public access to the raw data. We urge SILSO management to start the process to make this happen. A shining example of open data policy can be found in Scherrer at al. [2013].

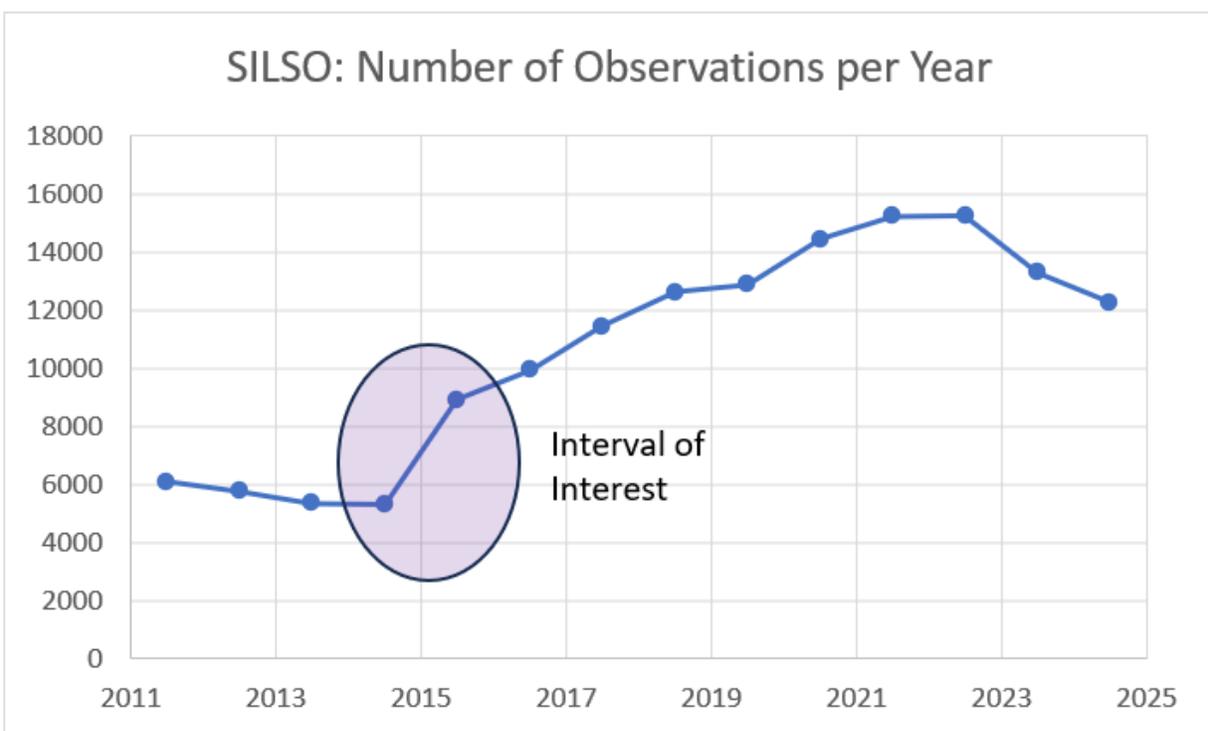

*Figure 8:* Tthe number of sunspot observations reported per year in the SILSO database (from https://www.sidc.be/SILSO/DATA/SN_y_tot_V2.0.txt)